\documentclass[twocolumn,showpacs,preprintnumbers,amsmath,amssymb]{revtex4}
\usepackage{color}
\usepackage{graphicx} 
\usepackage{dcolumn} 

\newcommand{\etal}{\it et al.}

\newcommand{\bto}{BaTiO$_3$}

\newcommand{\half}{\frac{1}{2}}

\newcommand{\Ploc}{P_{\rm local}}

\begin{document}

\title{First-principles study of symmetry lowering in relaxed
  BaTiO$_3$/SrTiO$_3$ superlattices}

\author{Karen Johnston}
\email{karenjoh@physics.rutgers.edu}

\author{Xiangyang Huang}
\altaffiliation[Present address:]
{Department of Chemical Engineering and Materials Science
University of Minnesota, 
Minneapolis, MN 55455}

\author{Jeffrey B. Neaton}
\altaffiliation[Present address:]
{University of California at Berkeley,
Department of Physics,
Berkeley, California 94720}

\author{Karin M. Rabe}
\affiliation{Department of Physics and Astronomy, Rutgers
    University, Piscataway, NJ 08854}

\date{\today}

\begin{abstract}

The crystal structure and local spontaneous polarization of
(BaTiO$_3$)$_m$/(SrTiO$_3$)$_n$ superlattices is calculated using
a first-principles density functional theory method.  
The in-plane lattice constant is 1\% larger than the SrTiO$_3$
substrate to imitate the relaxed superlattice structure 
and the symmetry is lowered to monoclinic space group $Cm$ which
allows polarization to develop along the [110] and [001]
directions.  
The polarization component in the [110] direction is found to
develop only in the SrTiO$_3$ layers and falls to zero in
the BaTiO$_3$ layers, whereas the polarization in the [001]
direction is approximately uniform throughout the superlattice.
These findings are consistent with recent experimental data and
first-principles results for epitaxially strained BT and ST.    

\end{abstract}

\pacs{68.03.Hj, 68.65.Cd, 77.80.-e, 81.05.Zx}

\maketitle

\section{Introduction}

Perovskite materials, such as {\bto} (BT) and SrTiO$_3$ (ST), are
of considerable interest due to their polarization-related
properties. Short-period superlattices of these materials have
switchable polarization, ferroelectric T$_c$ and dielectric and
piezoelectric coefficients distinct from those of their
individual constituents, and the freedom to choose from a variety
of constituent materials, layer thicknesses and sequences can be
exploited to optimize selected desirable properties 
\cite{Tabata1994a,Sepliarsky2001a,Sepliarsky2001b,Christen2003a,
Jiang2003a,Shimuta2002a,Kim2002a,Neaton2003a,Tabata1997a}.  

The (BaTiO$_3$)$_m$-(SrTiO$_3$)$_n$ (BT/ST $m/n$) superlattice
has been extensively investigated by experimentalists primarily
because of the possible enhancement of the switchable
polarization arising from strain effects. Due to the lattice
mismatch with ST, there is a compressive in-plane strain on BT
that increases the tetragonal strain and, through
polarization-strain coupling, the [001] polarization.    
Several groups \cite{Tabata1994a,Tabata1995a,Shimuta2002a}
studied the structure and dielectric properties of symmetric and
asymmetric BT/ST superlattices with periods ranging from 4 to
500 unit cells.  Measurements of these superlattice structures
were consistent with tetragonal symmetry.  
Depending on the BT/ST ratio, the polarization can be enhanced
over that of unstrained BT and the Curie temperature elevated.  

Neaton and Rabe \cite{Neaton2003a} performed first-principles
calculations to study the structure and polarization
of BT/ST superlattices lattice-matched to an ST substrate. 
With this epitaxial constraint the systems studied ($m/n$
superlattices with $m+n=5$) were found to be stable with $P4mm$
symmetry, with polarization along [001].  
The total polarization was decomposed into contributions from 
individual Ti-centred layers.  It was found that the unstrained
ST layers were polarized along [001] and the strain-enhanced
polarization of BT reduced such that the polarization was nearly
constant in the [001] direction.  
This was attributed to the minimisation of electrostatic energy costs
associated with the buildup of polarization charge at the
interfaces, suggesting a simple model that provided a good fit to
the first-principles results.   

Recently Jiang {\etal} \cite{Jiang2003a} and R\'{\i}os {\etal}
\cite{Rios2003a} investigated the structure of BT/ST
superlattices with periods ranging from 4 to 100. Through
a combination of second harmonic generation and x-ray diffraction
measurements, they found that below a critical periodicity of
30/30, the system has symmetry lower than tetragonal, consistent
with a nonzero polarization in the ST layers along the [110]
direction.  
The in-plane lattice constants are $\approx$0.8\% larger than
that of bulk ST which means that the films are at least 
partially relaxed.
 
In this paper we describe first-principles calculations that
explore the idea that it is precisely this increase in in-plane
lattice constant that is responsible for the observed change in
structure.  For several BT/ST superlattices with period up to 10,
we expand the in-plane lattice constant to 1.01 times the
theoretical lattice constant of bulk ST and relax the tetragonal
symmetry constraint to allow polarization to develop along the
[110] direction as well as along [001].  We show that the ST
layers exhibit both the [001] polarization which, as in the
coherent tetragonal superlattices, minimises the electrostatic
energy due to charge accumulation at the BT-ST interface layers,
as well as a nonzero component of the polarization along [110],
lowering the symmetry as observed.
The details of the methods used are given in
Section~\ref{sec:method} and in Section~\ref{sec:results}
we present our results and discussion.  

\section{\label{sec:method}Method}

First-principles calculations were performed with the Vienna 
{\it ab initio} Simulations Package (VASP) \cite{Kresse1996a}
which implements density functional theory within the local
density approximation.  Projector augmented wave potentials were
used \cite{Blochl1994a,Kresse1999a} with the semicore Ba
($5s$,$5p$), Sr ($4s$,$4p$), Ti ($3s$,$3p$) treated as valence
states.  A plane wave energy cutoff of 600~eV and a
$6\times6\times2$ Monkhorst-Pack $k$-point sampling of the
Brillouin zone were used.  

The superlattices have structure $m/n$ where $m$ and $n$
represent the thickness in unit cells of BT and ST respectively.
The structures considered in this paper are 3/3, 4/4, 4/6, 5/5
and 6/4, with unit cells $1\times1\times(n+m)$.
The structures are optimized in the monoclinic space group $Cm$,
which allows nonzero polarization of the form
$P_1(\hat{x}+\hat{y}) + P_3\hat{z}$. 
Because the deviation from the in-plane square lattice observed
in the x-ray experiment is very small\cite{Rios2003a}, we
constrain the angles between the lattice vectors to 90
degrees.  The $c$ lattice parameter and the ions are allowed to
fully relax within the symmetry of the $Cm$ space group.  
The in-plane lattice parameter is fixed to
$a=1.01a_{\rm ST}=3.902$~{\AA}, with the calculated values of the
lattice constants for the constituent materials being 
$a_{\rm ST}=3.863$~{\AA} for bulk cubic ST and 
$a_{\rm BT}=3.943$~{\AA} and $c/a=1.012$ for bulk tetragonal 
BT.  This results in an epitaxial strain of -1.04\% on the
tetragonal BT layer.   
The relaxation was considered to be complete when the
Hellman-Feynman forces on the ions were less than
5~meV{\AA}$^{-1}$.

As discussed in the Introduction, this value of in-plane
lattice constant was chosen to mimic the experimental relaxation of the
superlattice.  This relaxation implies that the BT/ST
superlattices, although epitaxial, are not coherent with the
substrate and there must be dislocations present.  It is assumed
that these dislocations are confined to the region near the
substrate interface and therefore we are modelling the 
relaxed, defect-free superlattice far from the substrate
interface.

To investigate the spatial variation of the polarization and
strain within the superlattice, we compute the local polarization
in each 5-atom cell of the superlattices, defined as
\begin{eqnarray*}
P_{\rm local}
 &=& \frac{1}{\Omega}\{
     Z^*_{\rm Ti} \Delta z_{\rm Ti}
  +  Z^*_{\rm O1} \Delta z_{\rm O1}
  +  Z^*_{\rm O2} \Delta z_{\rm O2} \\
 &+& \half(
     Z^*_{\rm O3_{-}}\Delta z_{\rm O3_{-}}
  +  Z^*_{\rm O3_{+}}\Delta z_{\rm O3_{+}})\}
\end{eqnarray*}
where $Z^*_{\rm i}$ is the dynamical charge of atom $i$ 
\footnote[1]{ 
The dynamical charges are approximated by those of bulk
cubic BT in the BT layers, bulk cubic ST in the ST layers
and average values in the two TiO$_2$ interface layers.  
The values of the dynamical charges for cubic ST and tetragonal
BT were taken from Refs.~\onlinecite{Ghosez1998a,Zhong1994a}.  
The variation of $\Ploc$ depends on the choice of $Z^*$.
Even though the effect is small it should be noted that $\Ploc$
is only a guide to the variation of the polarization and should
not be treated as an exact quantity.  }
and $\Delta z_{\rm i}$ is the displacement in the $z$ direction
of atom $i$ with respect to the centrosymmetric cell defined by
the Ba/Sr atoms. 
${\rm O3_+}$ and ${\rm O3_-}$ represent the O3 atoms at the top
and bottom of the 5-atom cell, respectively.  
The local strain $c/a$ is obtained by using the A atoms to 
define the corners of the local unit cell, i.e.
$c/a = \left[z(\rm A_{+})-z(\rm A_{-})\right]/a$.

\section{\label{sec:results}Results and Discussion}

\begin{table}[ht!]
\caption{\label{tab:coa} Variation of $c/a$ in each
  superlattice.  The average $c/a$ ratios for the BT and ST
  layers are shown at the bottom of the table.  }
\begin{center}
\begin{tabular}{|r|rrrrr|} \hline
&\multicolumn{5}{|c|}{Present results} \\ 
  & 3/3  & 4/4  & 6/4  & 5/5  & 4/6 \\ \hline
\multicolumn{6}{|c|}{$c/a$ for 5 atom cells} \\ \hline
  &0.9999 &1.0000 &1.0003 &0.9999 &1.0000\\
  &0.984  &0.984  &0.986  &0.984  &0.983 \\
  &0.987  &0.987  &0.989  &0.987  &0.985 \\
  &1.010  &0.988  &0.990  &0.987  &0.985 \\
  &1.028  &1.010  &1.014  &0.988  &0.986 \\
  &1.026  &1.029  &1.031  &1.011  &0.985 \\   
  &    -  &1.027  &1.029  &1.029  &1.006 \\
  &    -  &1.027  &1.028  &1.027  &1.026 \\
  &    -  &    -  &1.028  &1.027  &1.024 \\
  &    -  &    -  &1.028  &1.027  &1.025 \\ \hline
\multicolumn{6}{|c|}{Average $c/a$}      \\ \hline
BT&1.027 &1.027 &1.029 &1.027 &1.025\\
ST&0.986 &0.986 &0.988 &0.987 &0.985\\
\hline
\end{tabular}
\end{center}
\end{table}

The local $c/a$ ratio for each layer is given in
Table~\ref{tab:coa}.  
The BT layers, being under compressive in-plane strain, have
$c/a$ larger than the bulk tetragonal value, whereas the ST 
layers are under tensile in-plane strain and have $c/a<1$.  The
two interface layers are inequivalent, with the polarisation
being directed towards the Ba in one and towards the Sr in the
other.  Both interfaces have $c/a$ values intermediate between
the average BT and ST values.  

\begin{figure}
\includegraphics[angle=-90,scale=0.45]{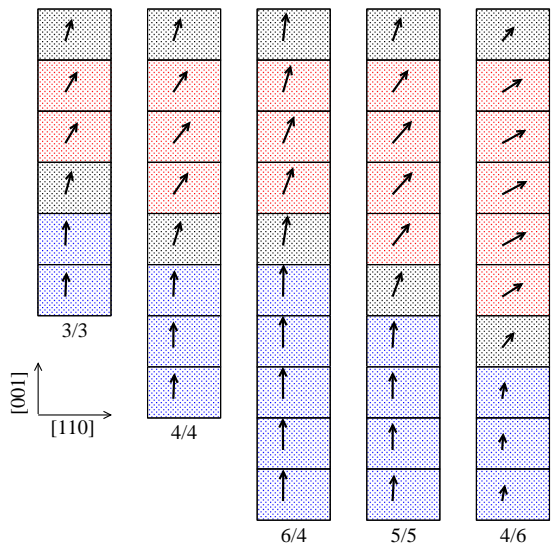}
\caption{\label{fig:plocal-vmap} (Color online) Magnitude and
  direction of $\Ploc$ in the various superlattices.  } 
\end{figure}

\begin{figure}
\includegraphics[angle=0,scale=0.6]{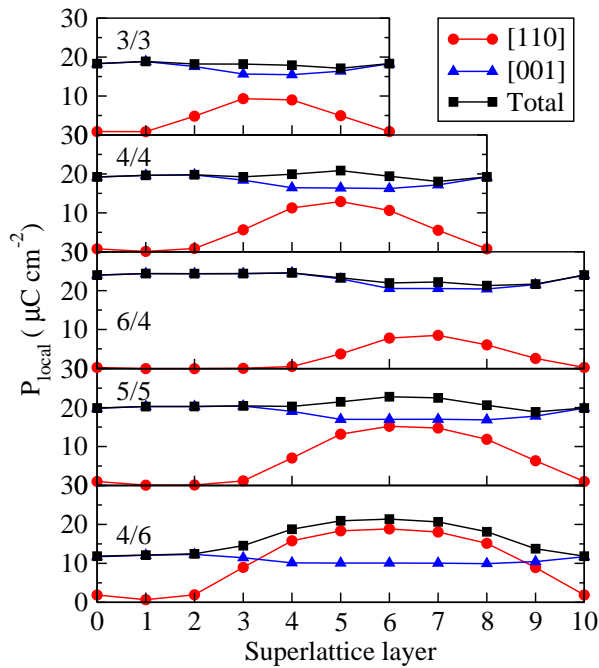}
\caption{\label{fig:plocal} (Color online) [001] and
  [110] components and magnitude of $\Ploc$.  }
\end{figure}

Figures~\ref{fig:plocal-vmap} and \ref{fig:plocal} show the
variation of $\Ploc$ in each superlattice.  
The [001] component remains approximately uniform throughout
the superlattice, dipping slightly in the ST layers.  
As the relative thickness of the BT layer increases, the
magnitude of the [001] polarization increases. 
In the ST layers, there is a nonzero polarization along [110],
lowering the symmetry from $P4mm$.  It grows smoothly from zero
at the interface over about two lattice constants, reaching its
maximum value in the center of the layer.  As the relative
thickness of the ST layer increases, the magnitude of the
polarization along [110] increases.  The [001] and [110]
components are not independent.  This can be seen by comparing
the 4/4 and 6/4 superlattices, in which the ST layer has the same
thickness. Relative to the 4/4 superlattice, in the 6/4
superlattice the [001] component is larger and the [110]
component is smaller.  
An analogous behavior is seen in the 4/4 and 4/6 superlattices.
In the 4/6 superlattice, the [001] component is smaller and the
[110] component is larger.  In all superlattices studied except
  the shortest period (3/3), the magnitude of the total
polarization in the center of the ST layer is quite similar.  

The local variation of polarization and strain can be understood
by applying the model discussed in Ref.~\onlinecite{Neaton2003a}
to the relaxed superlattice.  The first step is to consider the
effect of epitaxial strain on the polarization and $c/a$.  
In the relaxed BT/ST superlattice the BT layers are subject to
compressive in-plane strain whereas the ST layers are subject to
tensile in-plane strain.   
Di\'{e}guez {\etal} \cite{Dieguez2004a} studied the epitaxial
strain phase diagram of bulk BT in zero macroscopic field using a
first-principles approach.  They 
found that for compressive strains greater than 0.64\% BT is in
the $c$ phase (tetragonal structure $P4mm$), in which the
polarization points along the [001] direction.  
By extrapolating their structural parameters to the relaxed lattice
constant a=3.902~{\AA} we calculate the local polarization and
local $c/a$ of BT to be $28.4\mu$Ccm$^{-2}$ and 1.036, respectively.  
A similar study of the effect of strain on ST was carried out by
Antons {\etal} \cite{Antons2004a}.  They found that for 
tensile strains greater than 0.54\% ST has orthorhombic structure
$Amm2$ with polarization in the [110] direction.  The predicted
polarization and $c/a$ of bulk ST with an in-plane tensile strain
of 1\% are $\approx20\mu$Ccm$^{-2}$ and 0.985, respectively
\cite{Antons2004a}.

The next step is to take into account the electrostatic energy of
combining these two strained layers into a superlattice. In
particular, a discontinuity in the [001] component of the
polarization is energetically costly. The energy is reduced by
concomitant polarization of the highly polarizable ST layer and
reduction in the polarization of the BT layer to achieve
continuity. The final value of the polarization is strongly
dependent on the relative thicknesses of the two layers. The
thicker the BT layer relative to the ST layer, the higher the
polarization. These trends are evident in the computed
polarizations shown in Figure~\ref{fig:plocal}.

Through polarization-strain coupling, these changes in the
polarization lead to changes in the $c/a$ ratio.  This can at
least roughly account for the differences of the local BT and ST
$c/a$ from the strained bulk values, and for the trends between
$c/a$ and polarization in Table~\ref{tab:coa}. 

In the ST layer, increasing polarization along [001] reduces the
component along [110], minimizing changes in the total
magnitude.  This coupling has previously been noted in a
first-principles study \cite{Antons2004a} of phonon modes in
strained ST, where in the $Amm2$ phase, nonzero polarization
along [110] is associated with a hardening of the ferroelectric
mode along [001].  
This observation allows us to account for the observed variation
in the maximum [110] component in Figure~\ref{fig:plocal}.

In comparing these results with experiment, it is important to
select only systems with in-plane relaxation, such as the
superlattices in Ref.~\onlinecite{Rios2003a,Jiang2003a}.  
The shortest period system studied there is the 10/10
superlattice.  We can extrapolate the results from the three
symmetrical superlattices, namely the 3/3, 4/4 and 5/5.  The
$c/a$ values for each superlattice are remarkably similar so we
can expect the 10/10 superlattice to have similar $c/a$ values,
as indeed it does.  We can estimate the period of the 10/10
superlattice by scaling the periods of the 3/3, 4/4 and
5/5 superlattices.  This gives $\approx78.5$~{\AA} which
compares well with measured values of
78.1~{\AA}~\cite{Rios2003a} and 79~{\AA}~\cite{Shimuta2002a}. 
Consistent with the simple electrostatic model, the [001]
component increases only very slightly with increasing period,
and saturates to a estimated value of
$\approx22-23$~$\mu$Ccm$^2$, about 90\% percent of the
computed free bulk polarization of BT~\cite{Neaton2003a}.  
The [110] component at the centre of the ST layers increases from
less than 10~$\mu$Ccm$^{-2}$ for the 3/3 to
$\approx13\mu$Ccm$^{-2}$ for 4/4 to $\approx15\mu$Ccm$^{-2}$ for
the 5/5. This should be close to the saturation value; smaller
than the strained bulk ST polarization of
$\approx20\mu$Ccm$^{-2}$, as expected, due to the coupling to the
nonzero [001] component.  

To extrapolate to still longer periods, it is necessary to
recognize that the behavior will change once the thickness of the
layers exceeds the polarization coherence length and/or the
elastic critical thickness of the individual layers. The
uniformity of the [001] polarization is expected only for short
period superlattices, as for thicker layers the energy cost of
nonzero div P competes with the energy cost for polarizing the
paraelectric ST.  As a result, the [001] component decays into
the interior of the ST layers and for sufficiently thick layers
will become zero.  This behavior is demonstrated in
KNbO$_3$/KTaO$_3$ superlattices shell model studies by Sepliarsky
{\etal} \cite{Sepliarsky2001a,Sepliarsky2001b}.  In addition, as
the individual layers exceed their elastic critical thickness,
they would be expected to relax to their separate in-plane
lattice constants through the formation of misfit
dislocations. The precise value depends on the method of
synthesis; it has been shown that coherent films can be grown
well over the critical thickness under appropriate
conditions~\cite{Haeni2004a,Choi2004a}.

In summary it seems that the overall strain state is crucial to the
properties of the superlattice.  If the superlattice relaxes, as
in the work of Rios {\etal} the ferroelectric properties can be
dramatically different than in a superlattice where the strain is
preserved. For BT/ST, a simple model of electrostatics and bulk
strain effects works very well, with the very gentle interface
between BT and ST being relatively unimportant.

\begin{acknowledgments}
We would like to thank Refik Kortan, Susana R\'{\i}os, Andreas
R\"{u}diger and Jim Scott for valuable discussions.   This work
was supported by DOE Grant DE-FG02-01ER45937 and ONR Grant
N00014-00-1-0261.
\end{acknowledgments}

\bibliography{refs}

\begin{thebibliography}{10}

\bibitem{Tabata1994a}
H. Tabata, H. Tanaka, and T. Kawai, Appl. Phys. Lett. {\bf 65},  1970  (1994).

\bibitem{Sepliarsky2001a}
M. Sepliarsky {\it et~al.}, Phys. Rev. B {\bf 64},  060101  (2001).

\bibitem{Sepliarsky2001b}
M. Sepliarsky {\it et~al.}, Journal of Applied Physics {\bf 90},  4509  (2001).

\bibitem{Christen2003a}
H.~M. Christen, E.~D. Specht, S.~S. Silliman, and K. Harshavardhan, Phys. Rev.
  B {\bf 68},  020101(R)  (2003).

\bibitem{Jiang2003a}
A. Jiang, J. Scott, H. Lu, and Z. Chen, J. Appl. Phys. {\bf 93},  1180  (2003).

\bibitem{Shimuta2002a}
T. Shimuta {\it et~al.}, Journal of Applied Physics {\bf 91},  2290  (2002).

\bibitem{Kim2002a}
J. Kim {\it et~al.}, Integrated Ferroelectrics {\bf 47},  235  (2002).

\bibitem{Neaton2003a}
J. Neaton and K. Rabe, Appl. Phys. Lett. {\bf 82},  1586  (2003).

\bibitem{Tabata1997a}
H. Tabata and T. Kawai, Appl. Phys. Lett. {\bf 70},  321  (1997).

\bibitem{Tabata1995a}
H. Tabata, H. Tanaka, T. Kawai, and M. Okuyama, Jpn. J. Appl. Phys. I {\bf 34},
   544  (1995).

\bibitem{Rios2003a}
S. R\'{\i}os {\it et~al.}, J. Phys.: Condens. Matter {\bf 15},  L305  (2003),
  bST.

\bibitem{Kresse1996a}
G. Kresse and J. Furthm\"{u}ller, Phys. Rev. B {\bf 54},  11169  (1996).

\bibitem{Blochl1994a}
P. Bl\"{o}chl, Phys. Rev. B {\bf 50},  17953  (1994).

\bibitem{Kresse1999a}
G. Kresse and D. Joubert, Phys. Rev. B {\bf 59},  1758  (1999).

\bibitem{Dieguez2004a}
O. Di\'{e}guez {\it et~al.}, Phys. Rev. B {\bf 69},  212101  (2004).

\bibitem{Antons2004a}
A. Antons, J. Neaton, K.~M. Rabe, and D. Vanderbilt,
  http://arxiv.org/abs/cond-mat/0407077  (2004).

\bibitem{Haeni2004a}
J. Haeni {\it et~al.}, Nature {\bf 430},  758  (2004).

\bibitem{Choi2004a}
K. Choi {\it et~al.}, Science {\bf 306},  1005  (2004).

\bibitem{Ghosez1998a}
P. Ghosez, X. Gonze, and J.-P. Michenaud, Ferroelectrics {\bf 206-207},  205
  (1998).

\bibitem{Zhong1994a}
W. Zhong, R. King-Smith, and D. Vanderbilt, Phys. Rev. Lett. {\bf 72},  3618
  (1994).

\end{thebibliography}

\end{document}